\documentclass[lettersize,journal]{IEEEtran}
\IEEEoverridecommandlockouts

\usepackage{cite}
\usepackage{amsmath,amssymb,amsfonts}
\usepackage{algorithmic}
\usepackage{graphicx}
\usepackage{textcomp}
\usepackage{xcolor}
\usepackage{url}
\usepackage{subcaption}
\usepackage[caption=false,font=normalsize]{subfig}
\usepackage{siunitx}
\usepackage{threeparttable}

\graphicspath{{./figures/}}

\begin{document}

\title{Development of a Temperature-Variable Cold Blackbody Calibrator for Millimeter and Submillimeter Wave Detectors}

\author{
    \IEEEauthorblockN{
        Kazuki Watanabe,
        Taiki Sato,
        Shuhei Inoue,
        Shinsuke Uno,
        Tatsuya Takekoshi,
        and Tai Oshima}

\thanks{Received 26 September 2025; revised 28 May 2026; accepted 5 August 2026. Date of publication 10 August 2026; date of current version 24 August 2026. 
{\it (Corresponding author: Kazuki Watanabe.)}}
\thanks{Digital Object Identifier 10.1109/TASC.2026.3722206}
\thanks{Kazuki Watanabe and Tai Oshima are with National Astronomical Observatory of Japan (NAOJ), Tokyo 181-8588, Japan, and also with the Graduate University for Advanced Studies (SOKENDAI), Tokyo 181-8588, Japan.}
\thanks{Taiki Sato is with Saitama University, 255 Shimo-Okubo, Sakura-ku, Saitama 338-8570, Japan.}
\thanks{Shuhei Inoue is with the University of Tokyo, Tokyo 181-0015, Japan.}
\thanks{Shinsuke Uno is with RIKEN Center for Advanced Photonics, Saitama 351-0198, Japan.}
\thanks{Tatsuya Takekoshi is with Kitami Institute of Technology, Hokkaido 090-8507, Japan.}
}

\markboth{IEEE TRANSACTIONS ON APPLIED SUPERCONDUCTIVITY, VOL. 36, NO. 6, SEPTEMBER 2026}%
{Watanabe \MakeLowercase{\textit{et al.}}: Development of a Temperature-Variable Cold Blackbody Calibrator}

\makeatletter
\def\@IEEEpubidpullup{4.5\baselineskip}
\makeatother

\IEEEpubid{\fbox{\parbox{0.55\textwidth}{\footnotesize
© 2026 IEEE. Personal use of this material is permitted. Permission from IEEE must be obtained for all other uses, in any current or future media, including reprinting/republishing this material for advertising or promotional purposes, creating new collective works, for resale or redistribution to servers or lists, or reuse of any copyrighted component of this work in other works.}}}

\maketitle

\begin{abstract}
We report the design, fabrication, and characterization of a cryogenic blackbody calibrator {\color{black}intended for laboratory-based detector characterization, providing} multi-temperature calibration of millimeter- and submillimeter-wave detectors {\color{black}over a temperature range of 4~K to $\sim10$~K.} The device employs a Stycast/SiC absorber shaped for uniform array illumination and mounted on a phosphor-bronze substrate with optimized thermal links. Measured thermal conductance and heat capacity show the expected temperature dependence, although the conductance was lower than estimated, yielding a thermal time constant of 8.5~s at 10~K versus the 3.7~s design {\color{black}estimate}. The response remains practical, demonstrating the calibrator’s effectiveness as a stable reference source for precise calibration of detector arrays.
\end{abstract}


\section{Introduction}
\IEEEPARstart{W}{ide}-field observations in the millimeter and submillimeter regimes are key to probing cosmic evolution through studies of the cosmic microwave background (CMB), the hot intracluster plasma in galaxy clusters, and star formation in galaxies. These observations rely on highly sensitive cryogenic detectors, such as transition-edge sensors (TESs) and microwave kinetic inductance detectors (MKIDs).

For ground-based telescopes, however, thermal emission from the atmosphere dominates as the primary foreground. Its brightness temperature, typically 10--100~K depending on frequency and conditions, is more than three orders of magnitude higher than most astrophysical signals. To fully exploit detector sensitivity, accurate calibration of detector nonlinearity across a wide dynamic range is therefore essential.

The conventional two-point method, using blackbodies at room temperature (300~K) and liquid-nitrogen temperature (77~K), extrapolates detector response to lower temperatures under the assumption of linearity{\color{black}\cite{Ulich1976}}. This approach provides too few calibration points and is unreliable near 10~K, a critical regime that carries large statistical weight under favorable observing conditions.

\IEEEpubidadjcol

A cryogenic-free multi-temperature calibration source has been proposed to mitigate this issue. A motor-driven rotating filter wheel, mounted at the room-temperature section in front of the cryostat, holds eight graybody calibrators. This system delivers radiation at eight discrete temperatures spanning 25--100~K, matched to 270 and 350~GHz bands, and has been successfully demonstrated in the field \cite{Takekoshi2018, Oshima2018}. Despite its advantages, this approach faces two key challenges: the difficulty of realizing temperatures near 10~K required for lower-frequency observations, and the long driving time of the wheel, which forces a trade-off between calibration accuracy, efficiency, and the number of temperature points.

{\color{black}To address the need for cryogenic calibration, cold blackbody calibrators have been widely developed and are commonly used in CMB experiments~\cite{Henning2010, Choi2018, Anderson2020, Chuss2017, King2024}. These existing cold loads are optimized for high emissivity to serve as accurate absolute radiance references. However, they generally have a large heat capacity, which leads to long thermal time constants, making rapid multi-temperature sweeps difficult for efficient detector characterization, and also increases the thermal loading on the cryogenic environment surrounding the detectors.}

As an initial step toward overcoming these limitations, we developed a 
fast-response cold blackbody source for use in laboratory calibration experiments. The device is designed for installation on the 4~K stage of an observing system and provides variable-temperature operation {\color{black}from 4~K to 10~K}, enabling precise nonlinearity calibration of cryogenic detectors.

\section{Design Concept}
\label{sec:design_concept}
The cold blackbody developed in this study is intended for the calibration of submillimeter-wave detectors, targeting variable-temperature operation near 10~K.
For multi-temperature calibration, the blackbody temperature must be varied stepwise. A critical concern in this process is that thermal radiation from the blackbody can heat the detector wafer and the supporting sub-Kelvin stage. {\color{black}Since the cooling power of sub-Kelvin stages can be as low as $\sim$1~$\mu$W, particularly in systems employing sorption coolers, even a small additional radiative load can significantly affect the detector environment.} Therefore, minimizing the temperature rise of the detectors and surrounding cryogenic stages is an essential design requirement. To address this, the blackbody is mounted on the 4~K stage and positioned at a fixed distance in front of the cryogenic detectors.  

To suppress heating caused by thermal radiation from the blackbody, rapid on/off control of the emission is required, which demands a fast thermal response. From this perspective, we set a design goal of achieving a thermal time constant {\color{black}of less than 10~s}. To shorten the time constant, the heat capacity of the blackbody must be minimized while maximizing its thermal conductivity.  

However, achieving an emissivity close to unity requires a minimum thickness that depends on the observation frequency. This constraint imposes a lower limit on the size of the blackbody, which in turn restricts the illuminated area and thereby the number of detectors that can be calibrated simultaneously. Furthermore, while higher thermal conductivity enables faster cooling of the blackbody, it also increases heat flow into the 4~K stage, resulting in greater thermal loading on the cryogenic environment.  

Thus, in the design of the temperature-variable cold blackbody, there exists a fundamental trade-off between fast thermal response, sufficient illuminated area, and minimal thermal loading on the cryogenic system. In this study, the cold blackbody was designed with careful consideration of this trade-off.  

\subsection{Optical Coupling}
To deliver millimeter-wave radiation from the cold blackbody to the detector array, the following requirements were set:  
{\color{black}(1) the illuminated area should be as large as possible,}  
{\color{black}(2)} the radiation must uniformly illuminate the entire array, and  
{\color{black}(3)} stray light from external sources must be sufficiently suppressed so as not to compromise calibration accuracy.  

For optical coupling, the design is based on a conical horn array \cite{Takekoshi2012} and a TES bolometer array \cite{Oshima2013,Hirota2013}, which have been successfully implemented in the 270/350 GHz dual-color camera on the Atacama Submillimeter Telescope Experiment (ASTE). The parameters of the horn, such as aperture diameter and flare angle, are summarized in Table~\ref{tab:horn}. The horns are arranged in a hexagonal lattice with a spacing of 3.9~mm. {\color{black}Based on these parameters, the following requirements were established.}

\begin{table}[ht]
\centering
\caption{Horn parameters.}
\label{tab:horn}
\begin{tabular}{l r}
\hline
Parameter         & Value \\
\hline
Aperture diameter & 3.65~mm \\
Flare length      & 16.5~mm \\
Total flare angle & 12.7$^{\circ}$ \\
\hline
\end{tabular}
\end{table}

{\color{black}First, to suppress systematic errors arising from the optical design, we set a design target of an illumination edge taper of at least 10~dB, corresponding approximately to a beam leakage of 10\% or less. In the actual implementation, a band-defining optical filter must be placed between the horn aperture plane and the blackbody, which limits their minimum separation to 12.7~mm. Assuming Gaussian beam optics~\cite{goldsmith1998quasioptical} under this condition, the 10~dB attenuation beam radius $w_\mathrm{10dB}$ is 4.75~mm and 4.00~mm at 270~GHz and 350~GHz, respectively.}
 
{\color{black}Second, the target array is arranged in a regular hexagonal lattice with a pitch of 3.9~mm, and the maximum distance spanning the 19 pixels is $L = 3.9~\mathrm{mm} \times 4 = 15.6~\mathrm{mm}$. To ensure a taper of at least 10~dB for all pixels, the outer diameter $D$ of the blackbody was set as $D = L + 2\,w_\mathrm{10dB}$, which gives $D = 25.1~\mathrm{mm}$ at 270~GHz and $D = 23.6~\mathrm{mm}$ at 350~GHz.}
 
{\color{black}Finally, regarding the measurement accuracy,} the temperature was monitored with a Lakeshore silicon diode thermometer (DT-670A1-SD), which {\color{black}has} a calibration accuracy of $\pm 0.25$~K at 10~K. {\color{black}From the Rayleigh--Jeans approximation ($P \propto T$), the relative error in the radiated power is expressed as $\Delta P / P = \Delta T / T$ and amounts to about 2.5\% at 10~K. This error is sufficiently small compared with those arising from spillover and emissivity.}

\subsection{Thermal design}

\begin{table*}[htb!]
\centering
\caption{Estimated heat capacity of the cold blackbody at 10~K.}
\label{tab:heat_capacity}
\begin{threeparttable}
\begin{tabular}{lcccc}
\hline
Material & Specific heat (J/g/K) & Mass (g) & Thickness (mm) & Heat capacity (J/K) \\
\hline
phosphor-bronze & $1.1 \times 10^{-3}$ & 0.4\tnote{\textdagger} & 0.1 & $4 \times 10^{-4}$ \\
Stycast~2850FT  & $7.4 \times 10^{-3}$ & 0.8 & 0.9\tnote{\textdagger} & $6 \times 10^{-3}$ \\
SiC             & $1.0 \times 10^{-4}$ & 0.3 & 0.2\tnote{\textdagger} & $3 \times 10^{-5}$ \\
\hline
Total &  &  &  & $6 \times 10^{-3}$ \\
\hline
\end{tabular}
\begin{tablenotes}
\item[\textdagger] Estimated from thickness or mass.
\end{tablenotes}
\end{threeparttable}
\end{table*}

In designing the geometry of the blackbody, we first considered the cooling capacity of the refrigerator. The refrigerator used provides a cooling power of 0.6~W, and in order to suppress the temperature rise of the 4~K stage, the maximum heating power of the blackbody heater was limited to 10~mW.  
Under these conditions, the required thermal conductance to maintain the blackbody at 10~K is $G \lesssim 1.7$~mW/K. Furthermore, {\color{black}the heat capacity must be kept small for a fast response; for reference, a thermal time constant as short as $\tau = C/G \sim 1$~s would require $C \lesssim 1.7$~mJ/K.}

Next, we estimated the heat capacity of the blackbody. Following previous studies, the blackbody was fabricated from a composite material of Stycast~2850FT and SiC grains {\color{black}(Shinano Rundum C F36, grain size 500~$\mu$m)} \cite{Klaassen2001}. The thickness was set to 1~mm {\color{black}from the standpoint of reducing the heat capacity and shortening the thermal time constant.} {\color{black}The emissivity of the absorber was evaluated at room temperature by comparing the load curves of a TES bolometer, using the following references: an open cryostat window optically coupled to the ambient room-temperature environment ($\varepsilon \sim 1.0$), and a cryostat window covered with an aluminum plate ($\varepsilon \sim 0.03$); the emissivity of the 1~mm-thick blackbody was estimated to be $\varepsilon \approx 0.85$.} Using literature values of the specific heats of Stycast~2850FT and SiC at 10~K \cite{Javorsky2005, Lin1987}, together with a mixing ratio of 7:3 obtained in preliminary tests, the average specific heat was estimated to be 12~kJ/K/m$^{3}$. Consequently, the allowable blackbody area is limited to 142~mm$^2$.

However, such a small area would accommodate only 7 horns. Therefore, we adopted a hexagonal shape with an outer diameter of 25~mm, matching the hexagonal horn arrangement. In this configuration, 19 horns can be illuminated uniformly with an edge taper better than 10~dB. The heat capacity is about 6~mJ/K, and with the assumed thermal conductance, the resulting thermal time constant is about 3.5~s. {\color{black}This value satisfies the design goal of a time constant of less than 10~s, and a sufficient margin is retained even when the spread in thermal conductivity and specific heat among the individual materials is taken into account.}

We then designed the thermal link connecting the blackbody to the thermal bath. A 0.1~mm thick phosphor-bronze plate was chosen because it offers both moderate thermal conductivity and ease of fabrication. An additional advantage is that low-purity copper alloys generally show relatively small variations in material properties due to compositional differences. As shown in Fig.~1, two leg-like structures were extended from two sides of the hexagonal blackbody, each with a width of 12.5~mm and a length of 5~mm. This yields $G \approx 1.7$~mW/K, based on the thermal conductivity of phosphor-bronze \cite{Childs1973}. Furthermore, since the specific heat of phosphor-bronze is smaller than that of the blackbody material \cite{Simon1992}, it was also used as the substrate on which the blackbody was coated, thereby facilitating faster internal thermalization.  
The estimated heat capacities of the individual components and the total blackbody are summarized in Table~\ref{tab:heat_capacity}. The overall thermal time constant of this configuration is estimated to be 3.7~s.  

Finally, we verified that the internal thermalization of the blackbody would not limit the overall thermal response. Two-dimensional thermal diffusion can be approximated by $\tau_{\mathrm{intra}} = \frac{c}{\kappa} L^2$, where $c$ is the specific heat and $\kappa$ is the thermal conductivity. Taking the representative scale of the blackbody $L$ as 12.5~mm (half of the outer diameter of the hexagon), we obtain $\tau_{\mathrm{intra}} \approx 0.3$~s, indicating that thermal diffusion within the phosphor-bronze substrate is sufficiently fast. Moreover, the thermal transport from the phosphor-bronze to the blackbody surface, assuming the lower-limit thermal conductivity of Stycast~2850FT \cite{Tsai1978}, is estimated to be about 40~ms, which is still much shorter than the overall thermal time constant of the blackbody. Therefore, in this design, internal thermalization of the blackbody does not constitute a limiting factor in the response.

In summary, this thermal design fulfills the trade-off among heat capacity, thermal conductance, and illumination area identified earlier in this section.

\section{Fabrication and Evaluation of the Cold Blackbody}
\subsection{Fabrication}
\begin{figure}[htbp]
\centering
\includegraphics[width=\linewidth]{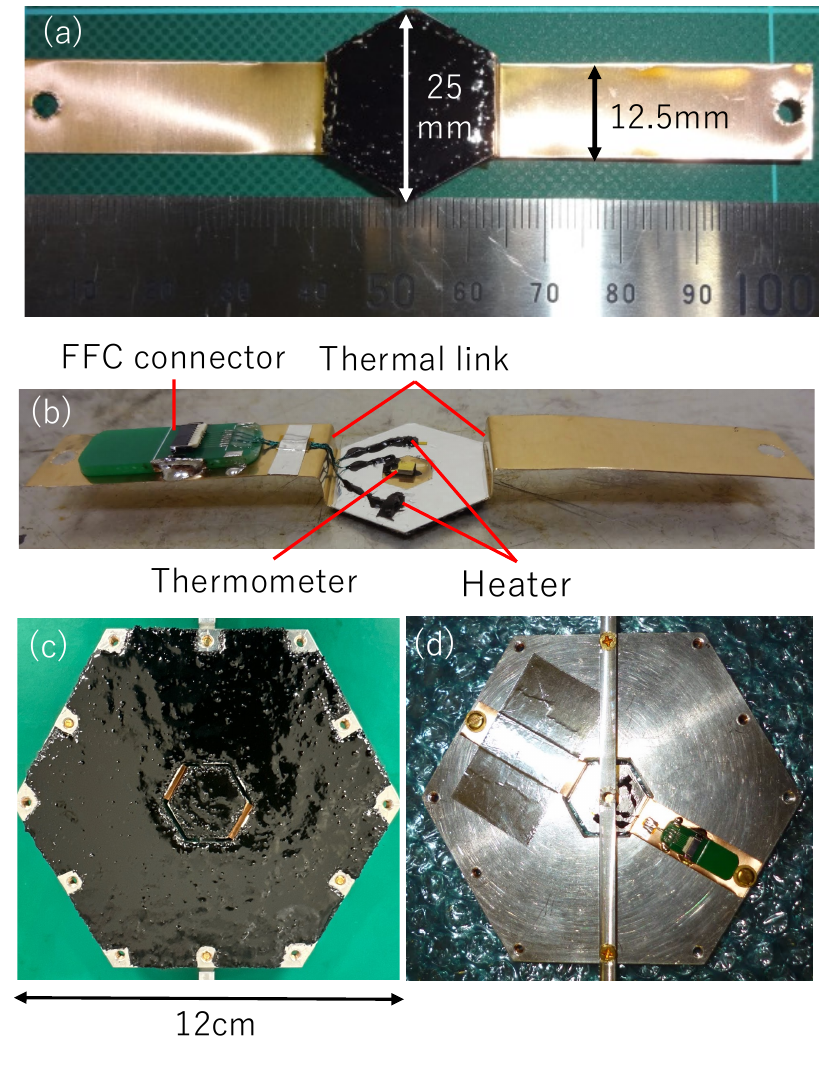}
\caption{Photographs of the fabricated cold blackbody: 
(a) blackbody surface side; 
(b) back side with heater and thermometer; 
(c) cold blackbody on a blackbody-coated holder plate (controlled background); 
(d) back side view of the cold blackbody with the holder plate.
}
\label{fig:bb}
\end{figure}

The structure of the fabricated cold blackbody is shown in Fig.~\ref{fig:bb}(a) and (b). A hexagonal absorber (thickness: 1~mm; outer diameter: 25~mm), composed of a mixture of Stycast~2850FT and SiC grains, was coated on a phosphor-bronze substrate. Two thermal links extend from opposite sides of the substrate and are bent in a crank shape toward the back side to block direct view from the detector side.

The back side of the substrate was covered with pure aluminum tape to reduce emissivity and thereby suppress radiative losses. A Lakeshore DT-670A1-SD thermometer was mounted on the top surface for temperature monitoring. Two 1~k$\Omega$ chip resistors (Panasonic ERJ-2GEJ102X) were connected in parallel to serve as a heater, ensuring efficient thermalization and redundancy in case of disconnection. These were affixed with a small amount of Stycast~2850FT while electrically insulated from the aluminum tape.

Electrical connections were made using three pairs of 0.1~mm-diameter constantan wires leading to an FFC (flat flexible cable) connector mounted on the holder. The thermal conductance of the wiring is approximately 10~$\mu$W/K, which is more than two orders of magnitude lower than that of the thermal links and therefore negligible.

The cold blackbody was mounted on a holder plate made of aluminum alloy (A5051), which was also coated with the same blackbody mixture. This configuration is intended to provide a controlled background that suppresses stray light from outside and maintains a stable background radiation level. A photograph of the mounted blackbody is shown in Fig.~\ref{fig:bb}(c) and (d).

\subsection{Measurement setup}
\begin{figure}[htbp]
\centering
\includegraphics[width=\linewidth]{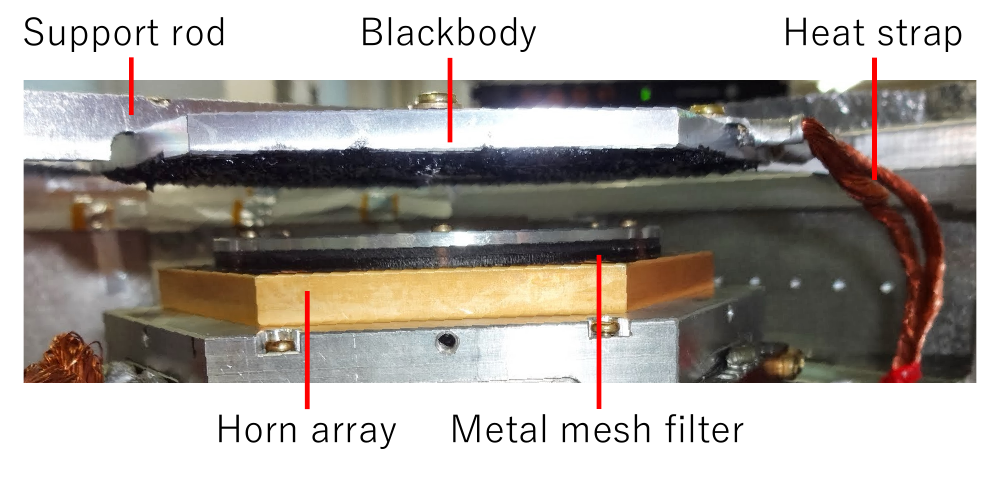}
\caption{A cold blackbody installed in the 4 K region of the cryostat, facing the horn array with a metal mesh filter placed between them. The blackbody is mounted on a holder plate suspended by an aluminum rod, with a 6N Cu heat strap attached to the holder plate. {\color{black}Note that, although a detector array was coupled to the horn array, it was not read out during this measurement, which was performed to evaluate the thermal influence of the blackbody on the horn array temperature.}}
\label{fig:bb_dewar}
\end{figure}

A photograph of the cold blackbody integrated into the 4~K stage of the cryostat is shown in Fig.~\ref{fig:bb_dewar}.  
The blackbody holder plate is thermally anchored to the 4~K stage, which serves as the heat bath, using two 6N copper wires (RRR $>$ 500) with an effective cross-sectional area of 2.7~mm${}^2$ and a length of 300~mm.  
In addition, the holder plate is suspended from the wall of the 4~K stage with an aluminum alloy (A5051) rod, thereby defining the distance between the cold blackbody and the horn array aperture plane as 12.5~mm. 

\subsection{Thermal characterization}
\begin{figure}[htbp]
\centering
\includegraphics[width=\linewidth]{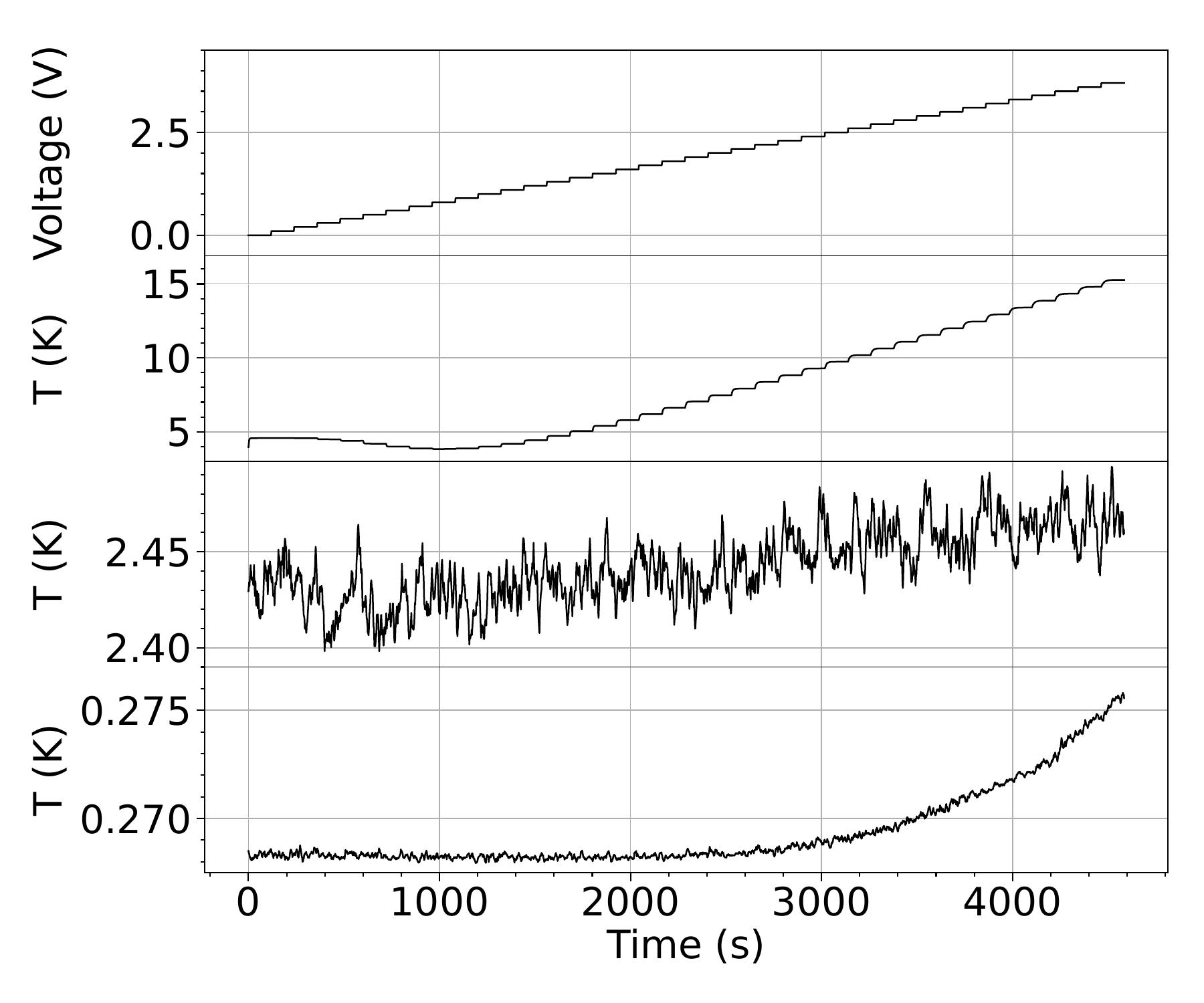}
\caption{Time response of the cold blackbody under stepwise heating. From top to bottom: heater control voltage, cold blackbody temperature, thermal bath temperature, and horn array temperature. (The minimum in the cold blackbody temperature trace at 0.8 V originates from a digital-to-analog converter (DAC) offset of 0.8 V.)}
\label{fig:bb_heatstep}
\end{figure}

To quantitatively evaluate the thermal response of the cold blackbody, we measured the temperature rise when the heater voltage was increased in 0.1~V steps every 120~s. As shown in Fig.~\ref{fig:bb_heatstep}, the temperature rise of the horn array became evident once the blackbody temperature exceeded 10~K. {\color{black}At $T_\mathrm{BB} = 10$~K, the temperature rise of the horn array is kept below 1~mK.} At the same time, the temperature of the thermal bath (4~K stage) also began to increase. These results confirm that the blackbody can be operated up to $\sim$10~K while keeping the thermal load on the surroundings sufficiently low.

\begin{figure}[htbp]
\centering
\includegraphics[width=\linewidth]{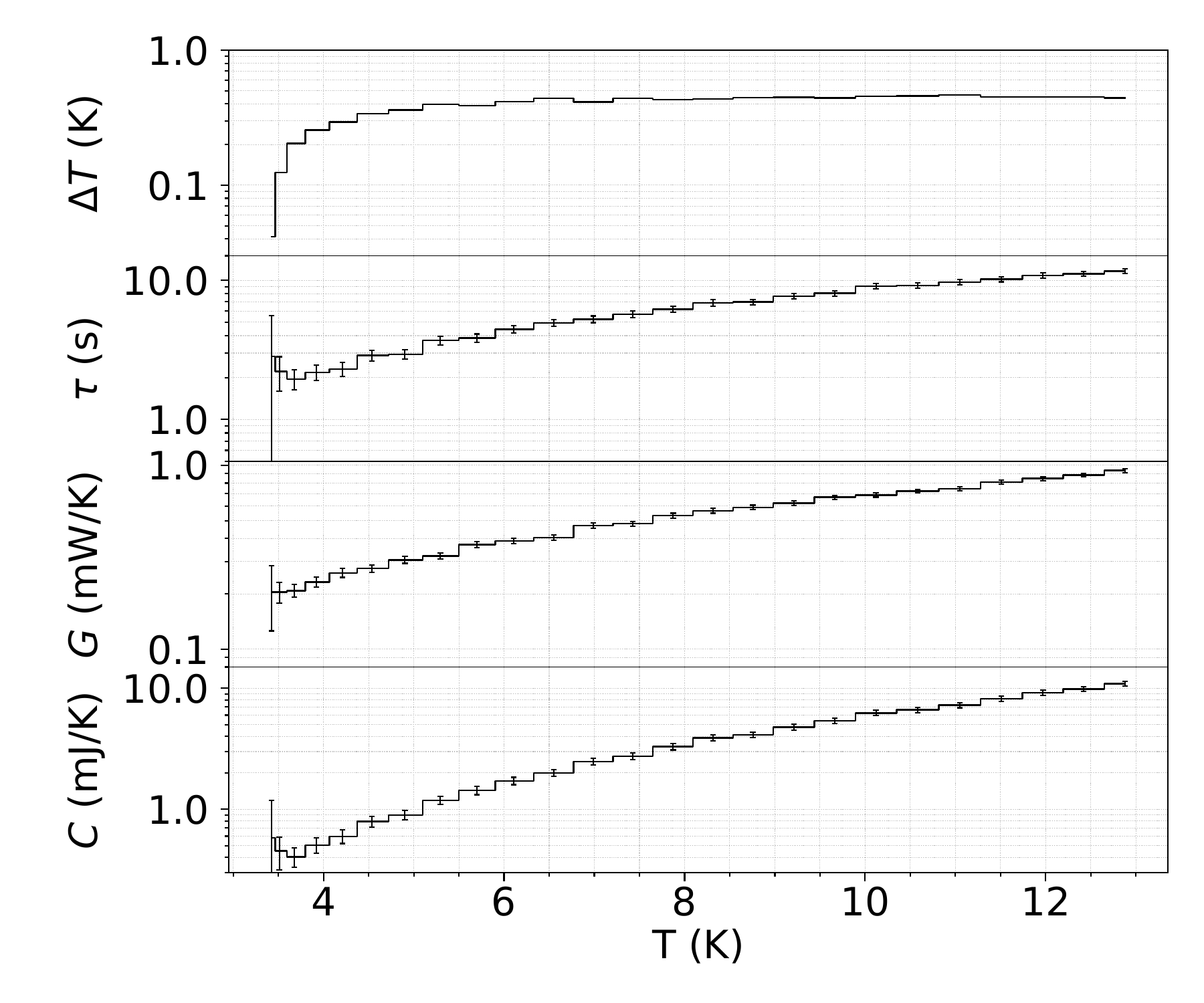}
\caption{Extracted thermal parameters of the cold blackbody. From top to bottom: cold blackbody temperature rise, thermal time constant, thermal conductance, and heat capacity.}
\label{fig:tau}
\end{figure}

From this measurement, the thermal conductance $G$, thermal time constant $\tau$, and heat capacity $C$ were derived.  
The average thermal conductance was calculated as  
$G(T) = \Delta P / \Delta T$,  
where $\Delta P$ and $\Delta T$ are the changes in power and temperature, respectively,  
and the representative temperature $T$ was defined as the average of the temperatures before and after the applied voltage change.  
The temporal temperature response was fitted with an exponential function to obtain the thermal time constant $\tau(T)$, and the heat capacity was then evaluated from  
$C(T) = \tau(T) \cdot G(T)$.  
The thermal parameters obtained at each temperature are shown in Fig.~\ref{fig:tau}.  

\begin{table}[htbp]
\centering
\caption{Estimated and measured values of thermal conductance, heat capacity, and thermal time constant of the cold blackbody.}
\label{tab:thermal_params}
\begin{tabular}{lccc}
\hline
 & $G$ (mW/K) & $C$ (mJ/K) & $\tau$ (s) \\
\hline
Estimated at 10~K & 1.7 & 6.3 & 3.7 \\
Measured at ~10~K & 0.7 & 5.8 & 8.5 \\
Temperature dependence & $0.048~T^{1.2}$ & $0.017~T^{2.5}$ & $0.36~T^{1.4}$ \\
\hline
\end{tabular}
\end{table}

The temperature dependences of the extracted parameters were well reproduced by the power-law functions listed in Table~\ref{tab:thermal_params}.  
The exponent of the thermal conductance (1.2) is close to the linear $T$ dependence typical of metallic materials below 10~K, indicating that the behavior of the phosphor-bronze thermal links was observed.  
The exponent of the heat capacity (2.5) is consistent with the reported value for Stycast~2850FT (2.2) \cite{Javorsky2005}.  
In contrast, comparison of absolute values at 10~K shows that the heat capacity agreed well with the estimate, while the thermal conductance was lower, likely due to uncertainties in the composition of the phosphor-bronze.  
As a result, the thermal time constant was somewhat longer (8.5~s) than the design value of $\approx 3.7$~s, but the response remains sufficiently practical.  

\begin{figure}[htbp]
\centering
\includegraphics[width=\linewidth]{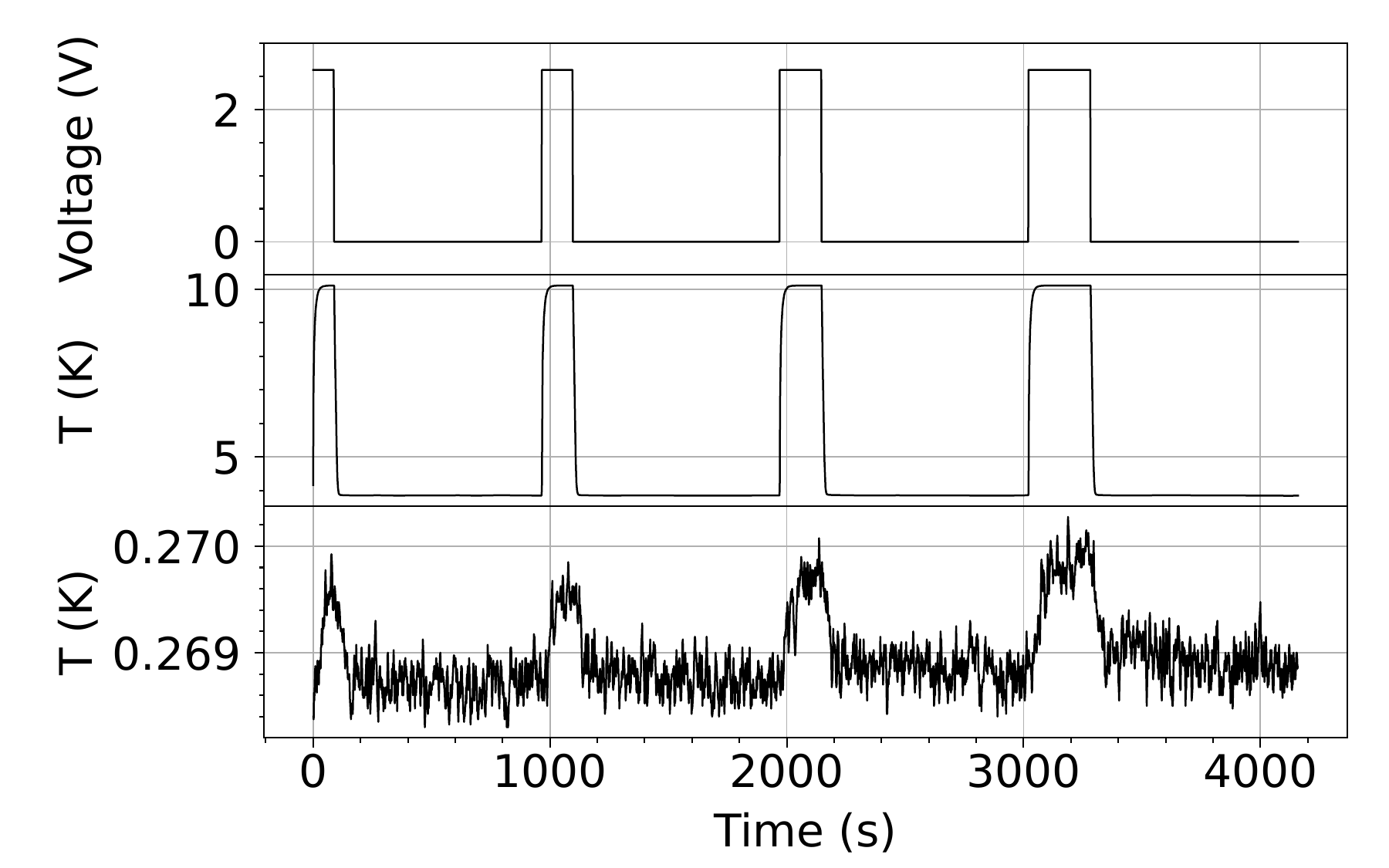}
\caption{{\color{black}Time response to heater pulse excitation with durations of 90, 120, 180, and 240~s.} From top to bottom: applied voltage, cold blackbody temperature, and horn array temperature.}
\label{fig:bb_heatpulse}
\end{figure}

Next, to evaluate the practical thermal relaxation time as a calibration source, rectangular voltage pulses with different durations were applied to the heater, where the pulse amplitude was chosen to correspond to the equilibrium blackbody temperature of 10~K.
Figure~\ref{fig:bb_heatpulse} shows the time variation of the applied voltage and the temperatures at each monitored point.  
Relative to the final equilibrium temperature of the cold blackbody, the response settled within the calibration accuracy of the thermometer after 25~s and converged within 1~mK after 70~s.  
During this process, the horn array temperature rise was also limited to $\approx 1$~mK.  
These results confirm that the blackbody can be switched on and off within a sufficiently short time, while keeping the influence on the surrounding temperatures negligible.  
Moreover, the effective relaxation time can be further shortened by active heater control. {\color{black}A slight baseline drift is also seen in the horn array temperature in Fig.~\ref{fig:bb_heatpulse}. This arises from a gradual thermal load on the surrounding environment due to prolonged radiation from the blackbody. Since the blackbody is driven in a pulsed manner in actual operation, the effect of this drift on the calibration accuracy is negligible.}  

\subsection{Evaluation of the time constant using MKID}
Temperature measurements alone have limited sampling rates, making it difficult to directly evaluate the fast thermal response of the cold blackbody. Therefore, in this study, we employed an MKID, which provides fast response and high sampling rates, to measure the rapid response characteristics of the blackbody.  

In the experiment, the cold blackbody was coupled to the MKID, and three rectangular voltage pulses, each of approximately 1~s duration but with different amplitudes, were applied to the blackbody heater. The corresponding responses were simultaneously recorded in terms of the MKID phase signal and the blackbody temperature $T_\mathrm{BB}$. The MKID signal was acquired at 1~ksps, clearly capturing both the sharp rise immediately after heating and the subsequent gradual decay. In contrast, the $T_\mathrm{BB}$ data had a sampling interval of about 3~s, which was insufficient to resolve the exact timing of rapid changes or peak temperatures, but the overall behavior agreed well with the MKID response, confirming its validity as an indicator of the thermal response.  

\begin{figure}[htbp]
\centering
\includegraphics[width=\linewidth]{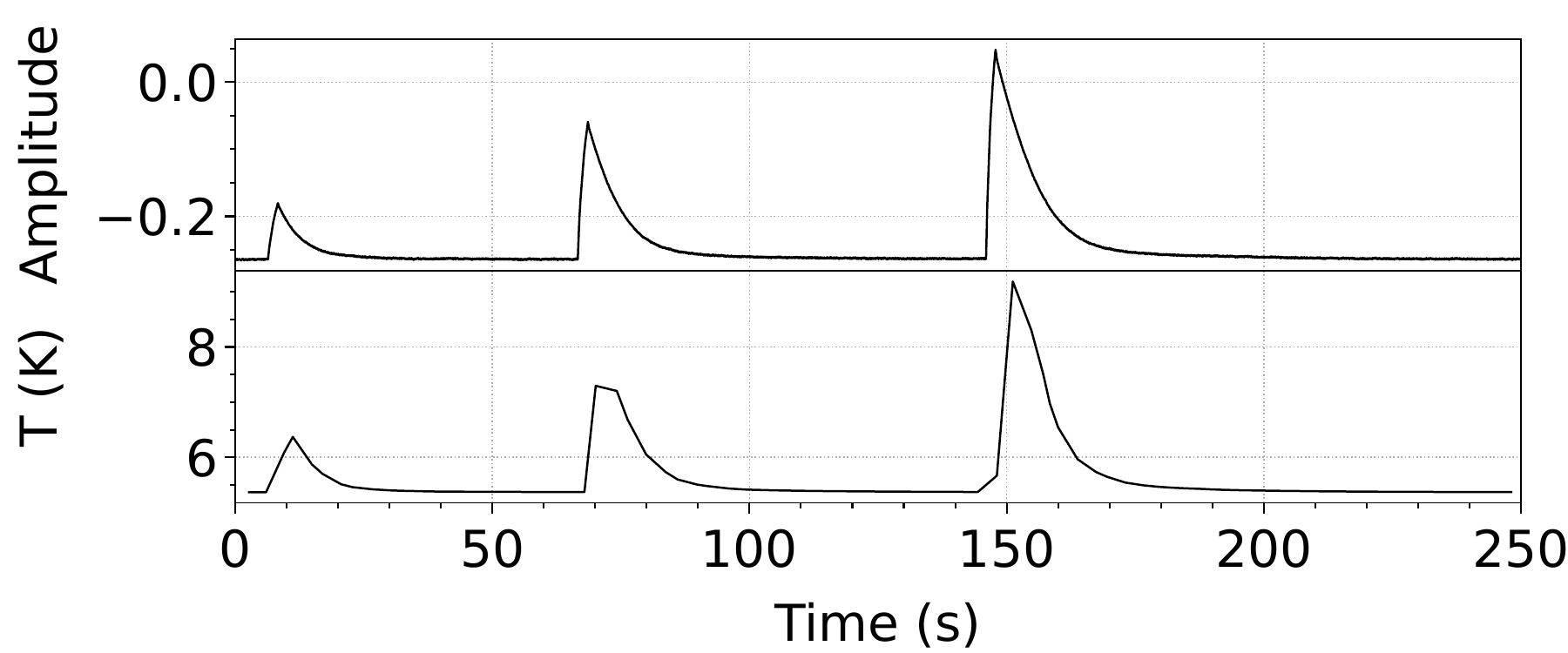}
\caption{Simultaneous measurement of MKID response and cold blackbody temperature. 
Top: Heating response observed in the MKID phase signal, showing a sharp rise and temperature-dependent variations in the decay. 
Bottom: Temporal variation of the cold blackbody temperature $T_\mathrm{BB}$ acquired at the same time. The three pulse inputs cause stepwise temperature increases, which correspond to features seen in the MKID response.}
\label{fig:mkid}
\end{figure}

Figure~\ref{fig:mkid} shows the recorded MKID response and $T_\mathrm{BB}$. The MKID phase signal exhibits a steep rise and exponential decay, which can be well represented by a two-component time-constant model. For each response, the rise ($\tau_0$) and decay ($\tau_1$) time constants were obtained through model fitting.  
The results are summarized in Table~\ref{tab:mkid_tau}. The value of $\tau_0$ remained nearly constant across all pulses and showed no temperature dependence. This suggests that $\tau_0$ reflects the thermalization time within the blackbody disk, rather than the intrinsic time constant of the MKID itself. In contrast, $\tau_1$ increased with rising blackbody temperature, reflecting the temperature dependence of the heat capacity and thermal conductance. This trend is consistent with the thermal response evaluation of the standalone blackbody, demonstrating that the MKID-based measurement successfully captures the fast thermal response characteristics of the cold blackbody.  

\begin{table}[htbp]
\centering
\caption{Rise ($\tau_0$) and decay ($\tau_1$) time constants derived from the MKID response.}
\label{tab:mkid_tau}
\begin{tabular}{lcc}
\hline
$T_\mathrm{BB}$ (K) & $\tau_0$ (s) & $\tau_1$ (s) \\
\hline
6.4 & 0.87 & 4.3 \\
7.4 & 0.95 & 5.6 \\
9.2 & 0.89 & 7.0 \\
\hline
\end{tabular}
\end{table}

\section{Conclusion}
We have developed and evaluated a cold blackbody calibrator designed for multi-temperature calibration of millimeter- and submillimeter-wave detectors {\color{black}over the temperature range of 4~K to 10~K}.  
A primary design objective was to satisfy the inherent trade-off among heat capacity, thermal conductance, and illumination area, enabling a calibrator with sufficiently fast thermal response while maintaining a practical illumination coverage for detector arrays.
The developed system successfully fulfills this trade-off, achieving {\color{black}a thermal time constant of less than 10~s}, consistent with the requirements derived from the design concept.  
Thermal properties were quantitatively characterized, showing good agreement with expectations for the constituent materials, though the measured conductance was slightly lower, leading to a somewhat longer time constant than designed.  
Experiments with MKIDs further verified the sub-second thermalization behavior within the blackbody disk, confirming that the calibrator exhibits the intended fast thermal response.  
These results demonstrate that the developed system provides a practical solution for accurate and efficient calibration of millimeter- and submillimeter-wave detector arrays in cryogenic environments.

\section*{Acknowledgements}
We thank the anonymous reviewers for their careful review and valuable comments.
This study was carried out in cooperation with the Advanced Technology Center of the National Astronomical Observatory of Japan (NAOJ).
K.~W. was supported by JST SPRING, Japan Grant Number JPMJSP2104.
S.~I. was supported by FoPM, WINGS Program, the University of Tokyo.
T.~T. was supported by the MEXT Leading Initiative for Excellent Young Researchers (Grant No. JPMXS0320200188).
This work was supported in part by RIKEN Special Postdoctoral Researcher Program,
JSPS KAKENHI Grant Numbers JP25KJ0981, JP24K22911, JP24H00004, JP23K25905, JP23K25879, JP23K20035, JP23H00121{\color{black},} 
the Murata Science and Education Foundation, the Nakajima Foundation, and the Sumitomo Foundation (Basic Science Research Grant No. 2200541).

\bibliographystyle{IEEEtran}
\bibliography{references}

@article{Ulich1976,
  author    = {B. L. Ulich and R. W. Haas},
  title     = {{Absolute Calibration of Millimeter-Wavelength Spectral Lines}},
  journal   = {Astrophys. J. Suppl. Ser.},
  volume    = {30},
  pages     = {247--258},
  year      = {1976},
  doi       = {10.1086/190361}
}

@article{Takekoshi2018,
  author    = {T. Takekoshi and K. Ohtawara and T. Oshima and S. Ishii and N. Izumi and T. Izumi and M. Yamaguchi and S. Suzuki and K. Muraoka and A. Hirota and F. Saito and S. Nakatsubo and A. Kouchi and T. Ito and K. Uemizu and Y. Fujii and Y. Tamura and K. Kohno and R. Kawabe},
  title     = {{Development of Multi-temperature Calibrator for the TES Bolometer Camera: System Design}},
  journal   = {J. Low Temp. Phys.},
  volume    = {193},
  number    = {5--6},
  pages     = {1003--1009},
  year      = {2018},
  doi       = {10.1007/s10909-018-1916-1}
}

@article{Oshima2018,
  author    = {T. Oshima and K. Ohtawara and T. Takekoshi and S. Ishii and N. Izumi and T. Izumi and M. Yamaguchi and S. Suzuki and K. Muraoka and A. Hirota and F. Saito and S. Nakatsubo and A. Kouchi and T. Ito and K. Uemizu and Y. Fujii and Y. Tamura and K. Kohno and R. Kawabe},
  title     = {{Development of Multi-temperature Calibrator for the TES Bolometer Camera: Deployment at ASTE}},
  journal   = {J. Low Temp. Phys.},
  volume    = {193},
  number    = {5--6},
  pages     = {996--1002},
  year      = {2018},
  doi       = {10.1007/s10909-018-2009-x}
}

@inproceedings{Henning2010,
  author    = {J. W. Henning and J. W. Appel and J. E. Austermann and J. A. Beall and D. Becker and D. A. Bennett and L. E. Bleem and B. A. Benson and J. Britton and J. E. Carlstrom and C. L. Chang and H. M. Cho and A. T. Crites and T. Essinger-Hileman and W. Everett and E. M. George and N. W. Halverson and G. C. Hilton and W. L. Holzapfel and J. Hubmayr and K. D. Irwin and D. Li and J. McMahon and J. Mehl and S. S. Meyer and S. Moseley and J. P. Nibarger and M. D. Niemack and L. P. Parker and E. Shirokoff and S. M. Simon and S. T. Staggs and J. N. Ullom and K. U-Yen and C. Visnjic and E. Wollack and K. W. Yoon and E. Y. Young and Y. Zhao},
  title     = {{Optical efficiency of feedhorn-coupled TES polarimeters for next-generation CMB instruments}},
  booktitle = {Millimeter, Submillimeter, and Far-Infrared Detectors and Instrumentation for Astronomy V},
  series    = {Proc. SPIE},
  volume    = {7741},
  pages     = {774122},
  year      = {2010},
  doi       = {10.1117/12.859478}
}

@article{Choi2018,
  author    = {S. K. Choi and J. Austermann and J. A. Beall and K. T. Crowley and R. Datta and S. M. Duff and P. A. Gallardo and S. P. Ho and J. Hubmayr and B. J. Koopman and Y. Li and F. Nati and M. D. Niemack and L. A. Page and M. Salatino and S. M. Simon and S. T. Staggs and J. Stevens and J. Ullom and E. J. Wollack},
  title     = {{Characterization of the Mid-Frequency Arrays for Advanced ACTPol}},
  journal   = {J. Low Temp. Phys.},
  volume    = {193},
  number    = {3--4},
  pages     = {267--275},
  year      = {2018},
  doi       = {10.1007/s10909-018-1982-4}
}

@article{Anderson2020,
  author    = {A. J. Anderson and P. A. R. Ade and Z. Ahmed and J. S. Avva and P. S. Barry and R. {Basu Thakur} and A. N. Bender and B. A. Benson and L. Bryant and K. Byrum and J. E. Carlstrom and F. W. Carter and T. W. Cecil and C. L. Chang and H.-M. Cho and J. F. Cliche and A. Cukierman and T. de Haan and E. V. Denison and J. Ding and M. A. Dobbs and D. Dutcher and W. Everett and K. R. Ferguson and A. Foster and J. Fu and J. Gallicchio and A. E. Gambrel and R. W. Gardner and A. Gilbert and J. C. Groh and S. T. Guns and R. Guyser and N. W. Halverson and A. H. Harke-Hosemann and N. L. Harrington and J. W. Henning and G. C. Hilton and W. L. Holzapfel and D. Howe and N. Huang and K. D. Irwin and O. B. Jeong and M. Jonas and A. Jones and T. S. Khaire and A. M. Kofman and M. Korman and D. L. Kubik and S. Kuhlmann and C.-L. Kuo and A. T. Lee and E. M. Leitch and A. E. Lowitz and S. S. Meyer and D. Michalik and J. Montgomery and A. Nadolski and T. Natoli and H. Nguyen and G. I. Noble and V. Novosad and S. Padin and Z. Pan and P. Paschos and J. Pearson and C. M. Posada and W. Quan and A. Rahlin and D. Riebel and J. E. Ruhl and J. T. Sayre and E. Shirokoff and G. Smecher and J. A. Sobrin and A. A. Stark and J. Stephen and K. T. Story and A. Suzuki and K. L. Thompson and C. Tucker and L. R. Vale and K. Vanderlinde and J. D. Vieira and G. Wang and N. Whitehorn and V. Yefremenko and K. W. Yoon and M. R. Young},
  title     = {{Performance of Al--Mn Transition-Edge Sensor Bolometers in SPT-3G}},
  journal   = {J. Low Temp. Phys.},
  volume    = {199},
  number    = {1--2},
  pages     = {320--329},
  year      = {2020},
  doi       = {10.1007/s10909-019-02259-7}
}

@article{Chuss2017,
  author    = {D. T. Chuss and K. Rostem and E. J. Wollack and L. Berman and F. Colazo and M. DeGeorge and K. Helson and M. Sagliocca},
  title     = {{A cryogenic thermal source for detector array characterization}},
  journal   = {Rev. Sci. Instrum.},
  volume    = {88},
  number    = {10},
  pages     = {104501},
  year      = {2017},
  doi       = {10.1063/1.4996751}
}

@article{King2024,
  author    = {C. L. King and I. Gullett and A. J. Anderson and B. A. Benson and R. Bihary and H. Fan and J. M. Nagy and H. Nguyen and J. E. Ruhl and S. M. Simon},
  title     = {{Design and validation of a cold load for characterization of cosmic microwave background stage 4 detectors}},
  journal   = {J. Astron. Telesc. Instrum. Syst.},
  volume    = {10},
  number    = {4},
  pages     = {048003},
  year      = {2024},
  doi       = {10.1117/1.JATIS.10.4.048003}
}

@article{Takekoshi2012,
  author    = {T. Takekoshi and T. Minamidani and S. Nakatsubo and T. Oshima and M. Kawamura and H. Matsuo and T. Sato and N. W. Halverson and A. T. Lee and W. L. Holzapfel and Y. Tamura and A. Hirota and K. Suzuki and T. Izumi and K. Sorai and K. Kohno and R. Kawabe},
  title     = {{Optics Design and Optimizations of the Multi-Color TES Bolometer Camera for the ASTE Telescope}},
  journal   = {IEEE Trans. Terahertz Sci. Technol.},
  volume    = {2},
  number    = {6},
  pages     = {584--592},
  year      = {2012},
  doi       = {10.1109/TTHZ.2012.2218102}
}

@article{Oshima2013,
  author    = {T. Oshima and M. Kawamura and B. Westbrook and T. Sato and A. Suzuki and T. Takekoshi and K. Suzuki and T. Minamidani and A. Hirota and T. Izumi and A. T. Lee and W. Holzapfel and K. Kohno and R. Kawabe},
  title     = {{Development of TES Bolometer Camera for ASTE Telescope: I. Bolometer Design}},
  journal   = {IEEE Trans. Appl. Supercond.},
  volume    = {23},
  number    = {3},
  pages     = {2101004},
  year      = {2013},
  doi       = {10.1109/TASC.2013.2240751}
}

@article{Hirota2013,
  author    = {A. Hirota and B. Westbrook and K. Suzuki and T. Izumi and T. Takekoshi and T. Sato and T. Oshima and T. Minamidani and M. Kawamura and A. Suzuki and A. T. Lee and W. L. Holzapfel and K. Kohno and R. Kawabe},
  title     = {{Development of TES Bolometer Camera for ASTE Telescope: II. Performance of Detector Arrays}},
  journal   = {IEEE Trans. Appl. Supercond.},
  volume    = {23},
  number    = {3},
  pages     = {2101305},
  year      = {2013},
  doi       = {10.1109/TASC.2013.2247791}
}

@inproceedings{Klaassen2001,
  author    = {T. O. Klaassen and M. C. Diez and J. H. Blok and C. Smorenburg and K. J. Wildeman and G. Jakob},
  title     = {{Optical Characterization of Absorbing Coatings for Sub-millimeter Radiation}},
  booktitle = {Proc. 12th Int. Symp. Space Terahertz Technology},
  address   = {San Diego, CA, USA},
  pages     = {400--409},
  year      = {2001}
}

@article{Javorsky2005,
  author    = {P. Javorsk{\'y} and F. Wastin and E. Colineau and J. Rebizant and P. Boulet and G. Stewart},
  title     = {{Low-temperature heat capacity measurements on encapsulated transuranium samples}},
  journal   = {J. Nucl. Mater.},
  volume    = {344},
  number    = {1--3},
  pages     = {50--55},
  year      = {2005},
  doi       = {10.1016/j.jnucmat.2005.04.015}
}

@article{Lin1987,
  author    = {G. J. Lin and J. C. Ho and D. P. Dandekar},
  title     = {{Low-temperature heat capacities of silicon carbide}},
  journal   = {J. Appl. Phys.},
  volume    = {61},
  number    = {11},
  pages     = {5198--5201},
  year      = {1987},
  doi       = {10.1063/1.338302}
}

@techreport{Childs1973,
  author      = {G. E. Childs and L. J. Ericks and R. L. Powell},
  title       = {{Thermal Conductivity of Solids at Room Temperature and Below: A Review and Compilation of the Literature}},
  institution = {National Bureau of Standards},
  address     = {Boulder, CO, USA},
  type        = {NBS Monograph},
  number      = {131},
  year        = {1973},
  doi         = {10.6028/NBS.MONO.131}
}

@techreport{Simon1992,
  author      = {N. J. Simon and E. S. Drexler and R. P. Reed},
  title       = {{Properties of Copper and Copper Alloys at Cryogenic Temperatures}},
  institution = {National Institute of Standards and Technology},
  address     = {Boulder, CO, USA},
  type        = {NIST Monograph},
  number      = {177},
  year        = {1992},
  doi         = {10.6028/NIST.MONO.177}
}

@article{Tsai1978,
  author    = {C. L. Tsai and H. Weinstock and W. C. Overton},
  title     = {{Low temperature thermal conductivity of Stycast 2850FT}},
  journal   = {Cryogenics},
  volume    = {18},
  number    = {9},
  pages     = {562--563},
  year      = {1978},
  doi       = {10.1016/0011-2275(78)90162-5}
}

@book{goldsmith1998quasioptical,
  author    = {P. F. Goldsmith},
  title     = {{Quasioptical Systems: Gaussian Beam Quasioptical Propagation and Applications}},
  publisher = {IEEE Press},
  address   = {Piscataway, NJ, USA},
  year      = {1998},
  isbn      = {978-0-7803-3439-7}
}

\end{document}